\documentclass[fleqn,10pt]{wlscirep}
\usepackage[utf8]{inputenc}
\usepackage[T1]{fontenc}

\usepackage{amssymb}
\usepackage{physics}
\usepackage{float}
\usepackage{graphicx}
\usepackage{sidecap}
\usepackage{tabularx}
\usepackage{array}
\usepackage{booktabs}
\usepackage{tikz}
\usepackage{longtable}
\usepackage{hyperref}
\usepackage{booktabs}
\usepackage{cellspace} 
\usepackage{tabularx} 
\usepackage{float}

\title{\centering Low Crosstalk in a Scalable Superconducting Quantum Lattice}

\author[1, *]{\centering Mohammed Alghadeer} 

\author[1]{Shuxiang Cao}

\author[1]{Simone D Fasciati}

\author[1]{Michele Piscitelli}

\author[2]{\newline Paul C. Gow}

\author[2]{James C. Gates}

\author[1, *]{Mustafa Bakr} 

\author[1, *]{Peter J. Leek} 

\affil[1]{Department of Physics, Clarendon Laboratory, University of Oxford, Oxford, OX1 3PU, United Kingdom}

\affil[2]{Optoelectronics Research Centre, University of Southampton, Southampton, SO17 1BJ, United Kingdom}

\affil[*]{mohammed.alghadeer@physics.ox.ac.uk}
\affil[*]{mustafa.bakr@physics.ox.ac.uk}
\affil[*]{peter.leek@physics.ox.ac.uk}

\begin{abstract}
Superconducting quantum circuits are a key platform for advancing quantum information processing and simulation. Scaling efforts currently encounter challenges such as Josephson-junction fabrication yield, design frequency targeting, and crosstalk arising both from spurious microwave modes and intrinsic interactions between qubits. We demonstrate a scalable 4x4 square lattice with low crosstalk, comprising 16 fixed-frequency transmon qubits with nearest-neighbor capacitive coupling that is implemented in a tileable, 3D-integrated circuit architecture with off-chip inductive shunting to mitigate spurious enclosure modes. We report on the design and comprehensive characterization, and show that our implementation achieves targeted device parameters with very low frequency spreads and simultaneous single-qubit gate errors across the device. Our results provide a promising pathway toward a scalable, low-crosstalk superconducting lattice topology with high qubit connectivity for quantum error correction and simulation.

\end{abstract}

\begin{document}

\flushbottom
\maketitle

\section*{Introduction}

Realizing large-scale superconducting quantum circuits containing individually addressable, high-coherence qubits remains a significant hardware challenge toward utility-scale quantum computing \cite{kim2023evidence, preskill2023quantum}. Scalable two-dimensional (2D) lattice architectures enable the implementation of logical operations using quantum error-correction codes \cite{shor1995scheme, gottesman1997stabilizer, putterman2024hardware, brock2024quantum, lacroix2024scaling}, such as the surface code \cite{acharya2024quantum, eickbusch2024demonstrating}, and the simulation of 2D lattice Hamiltonian, including the Bose–Hubbard model \cite{karamlou2024probing}, in condensed matter \cite{rosenberg2024dynamics, cochran2024visualizing, gyawali2024observation, google2023measurement, malz2021topological} and atomic physics \cite{daley2022practical}. While increasing qubit counts is crucial for realizing practical applications, scaling superconducting qubits currently introduces significant obstacles related to maintaining overall device performance and integrating high connectivity without introducing additional errors in gate operations \cite{In13}. Although significant efforts have been made to improve fabrication techniques \cite{alghadeer2025characterization, alghadeer2024mitigating, alghadeer2022surface-30, chistolini2024performance, altoe2022localization-27} and minimize hardware requirements \cite{bakr2024multiplexed, bakr2024dynamic, fasciati2024complementing}, scaling superconducting qubits using simplified architectures is crucial to mitigate fabrication defects \cite{mohseni2024build, zanuz2024mitigating, abdurakhimov2022identification, bilmes2020resolving} and reduce hardware overhead \cite{In14}. Ultimately, a practical superconducting quantum computer must integrate a large number of physical qubits with robust connectivity and fast, high-fidelity gates for fault-tolerant protocols \cite{mohseni2024build}.

Scaling superconducting circuits requires coupling qubits while mitigating parasitic interactions \cite{le2023scalable}. Superconducting qubits, the most common type of which is the transmon \cite{transmon}, can be capacitively coupled via lithographically defined circuit elements \cite{place2021new}. These couplings must be carefully engineered to support universal two-qubit gates \cite{yan2018tunable}, without introducing additional crosstalk levels that scale up with qubit count. Crosstalk arising from unwanted interactions between qubits is related to residual inter-qubit coupling \cite{ketterer2023characterizing, tripathi2022suppression, murali2020software}. This coupling results in spatially correlated crosstalk which gives rise to an always-on, state-dependent ZZ shift between qubits \cite{krinner2020benchmarking, zhao2022quantum, fors2024comprehensive}. On one hand, if the ZZ coupling is sufficiently strong, it can be used for implementing entangling controlled-phase (CPHASE) or controlled-Z (CZ) gates \cite{chow2011simple, chow2013microwave-2, xu2020high-2, wei2022hamiltonian, xu2020high, collodo2020implementation, sung2021realization, stehlik2021tunable, long2021universal, chu2021coupler}; on the other hand, it constitutes an unwanted coherent error that degrades the fidelity of simultaneous single- and two-qubit gates and, by extension, quantum algorithms \cite{chen2023voltage, ganzhorn2020benchmarking, mckay2019three, bharti2022noisy}. These interactions give rise to correlated errors, violating the core assumption of independent error channels in quantum error correction, representing a critical obstacle to building fault-tolerant architectures~\cite{megrant2025scaling}. 

As the lateral dimensions of the chip increase, sections of metallization or ground planes can form low-frequency resonant modes in the electromagnetic (EM) environment, acting as parasitic coupling channels, that mediate additional unwanted qubit-qubit interactions \cite{kosen2024signal, das2024reworkable, huang2021microwave}. Even a moderate increase in the system size can give rise to parasitic paths that inadvertently couple to the qubits \cite{spring2020modeling}. One approach is to house qubits in 3D cavities whose fundamental frequencies lie well above the qubit transition energies, thus isolating the qubits from interfering with the cavity modes \cite{spring2022high}. Careful engineering of overall microwave environment (i.e., planar chip, its enclosure, control and readout channels) is crucial in engineering architectures \cite{krasnok2024superconducting}.

In this work, we present the design and experimental realization of a proof-of-concept, scalable 4×4 square lattice with fixed-frequency transmon qubits implemented in a tileable, 3D-integrated circuit quantum electrodynamics (cQED) architecture \cite{rahamim2017double, spring2020modeling, spring2022high, fasciati2024complementing}. By carefully engineering the device parameters and the microwave environment, we achieve very low frequency spread, crosstalk levels, and simultaneous single-qubit gate errors across the lattice without canceling always-on qubit-qubit coupling. We report on the detailed measurements of inter-qubit couplings, coherence times, single- and two-qubit gate errors. The simplicity of our engineering paves the way toward realizing a scalable superconducting lattice topology.

\section*{Methodology}
The Hamiltonian that describes a system of 16 transmon qubits in 4×4 lattice, in the anharmonic oscillator approximation, with fixed frequencies $\omega_i/2\pi$, anharmonicities $\alpha_i/2\pi$ and statically coupled by an exchange interaction $J_{i,j}$ between nearest-neighbour qubits, is expressed as \cite{dicarlo2009demonstration, chow2013microwave, magesan2020effective}

\begin{equation}
\frac{\hat H}{\hbar} = \sum_{i=0}^{15} \left( \omega_i + \frac{\alpha_i}{2} (\hat a_i^\dagger \hat a_i - 1) \right) \hat a_i^\dagger \hat a_i  
+
\sum_{\substack{i < j \\ j \in \mathcal{N}_i}} J_{i,j} (\hat{a}_i^\dagger \hat{a}_{j} + \hat{a}_i \hat{a}_{j}^\dagger) 
\label{eq.1}
\end{equation}

where the \(\mathcal{N}_i\) denotes the set of four nearest-neighbor qubits coupled to qubit \(i\). The ZZ shift, denoted by $\zeta$, quantifies how strongly the frequency of one qubit depends on the state of its neighboring qubits and normally has a significant additional contribution from higher-excited states \cite{krinner2020benchmarking, zhao2022quantum, fors2024comprehensive}. $\zeta$ can be defined in a two-qubit system by the energy difference

\begin{equation}
\zeta \;=\; E_{|\widetilde{11}\rangle} \;-\; E_{|\widetilde{10}\rangle} \;-\; E_{|\widetilde{01}\rangle} \;+\; E_{|\widetilde{00}\rangle},
\label{eq.2}
\end{equation}

where $E_{|\widetilde{i j}\rangle}$ denotes the energy of the dressed eigenstate $|\widetilde{i j}\rangle$, and the labels $i$ and $j$ here refer to the excitation number in each qubit. $\zeta$ is related to the exchange coupling $J_{ij}$ in Eq. \ref{eq.1} through the following expression \cite{solgun2019simple, solgun2022direct}

\begin{equation}
\zeta \approx -\frac{2J_{ij}^2(\alpha_i + \alpha_j)}{(\Delta_{ij} + \alpha_i)(\alpha_j - \Delta_{ij})},
\label{eq.3}
\end{equation}

where $\Delta_{ij} = \omega_i - \omega_j$ is the detuning between two qubits, with $J_{ij} \ll |\Delta_{ij}|$, and $\delta_i$ and $\delta_j$ are the two qubits anharmonicities. In practice, residual non-nearest-neighbour couplings can be mediated between any two qubits in the lattice through different mechanisms. This long-range interaction can arise due to higher-order virtual processes involving intermediate qubits and electromagnetic enclosure modes. We denote these non-nearest-neighbor couplings as \(\widetilde{J}_{i,j}\), which represent residual parasitic interactions. To account for these effects, the Hamiltonian is now extended to include additional terms that captures the contributions of \(\widetilde{J}_{i,j}\) between all non-nearest-neighbor qubits

\begin{equation}
\frac{\hat{H}}{\hbar} = \sum_{i=0}^{15} \left( \omega_i + \frac{\alpha_i}{2} (\hat{a}_i^\dagger \hat{a}_i - 1) \right) \hat{a}_i^\dagger \hat{a}_i 
+ 
\sum_{\substack{i < j \\ j \in \mathcal{N}_i}} J_{i,j} (\hat{a}_i^\dagger \hat{a}_{j} + \hat{a}_i \hat{a}_{j}^\dagger) 
+ 
\sum_{\substack{i < j \\ j \notin \mathcal{N}_i}} \widetilde{J}_{i,j} (\hat{a}_i^\dagger \hat{a}_j + \hat{a}_i \hat{a}_j^\dagger)
\label{eq.4}
\end{equation}

The Hamiltonian now includes both direct (nearest-neighbor) and indirect (long-range) couplings. In general, for any pair of qubits in the lattice, the exchange interaction $\mathcal{J}_{i,j}$ takes the form \cite{blais2021circuit}
\begin{equation}
    \mathcal{J}_{i,j} = \frac{2~E_{Ci}~E_{Cj}}{\hbar~E_{Cc}} 
    \left( \frac{E_{Ji}}{2 E_{Cj}} ~\frac{E_{Jj}}{2 E_{Cj}} \right)^{1/4},
    \label{eq:J_coupling}
\end{equation}

where $E_{J_i}$ and $E_{C_i}$ are the transmon Josephson and charging energies, and $E_{C_c} = \frac{e^2}{2C}$ is the charging energy of the fixed coupling capacitance $C$ between a pair of qubits. The coupling $\mathcal{J}$ here suggests that the interaction depends on frequencies of both qubits indirectly through the dependence on the Josephson energy $E_J$ of the transmons, since the Josephson energy $E_{Ji}$ is related to each qubit frequency by $\omega_{qi} \approx \frac{\sqrt{8 E_{Ji} E_{Ci}}}{\hbar}$, for a transmon qubit operating in the weakly anharmonic regime.


\subsubsection*{Device Architecture}
\begin{figure}[H]
  \centering
   \includegraphics[width=0.7\textwidth]{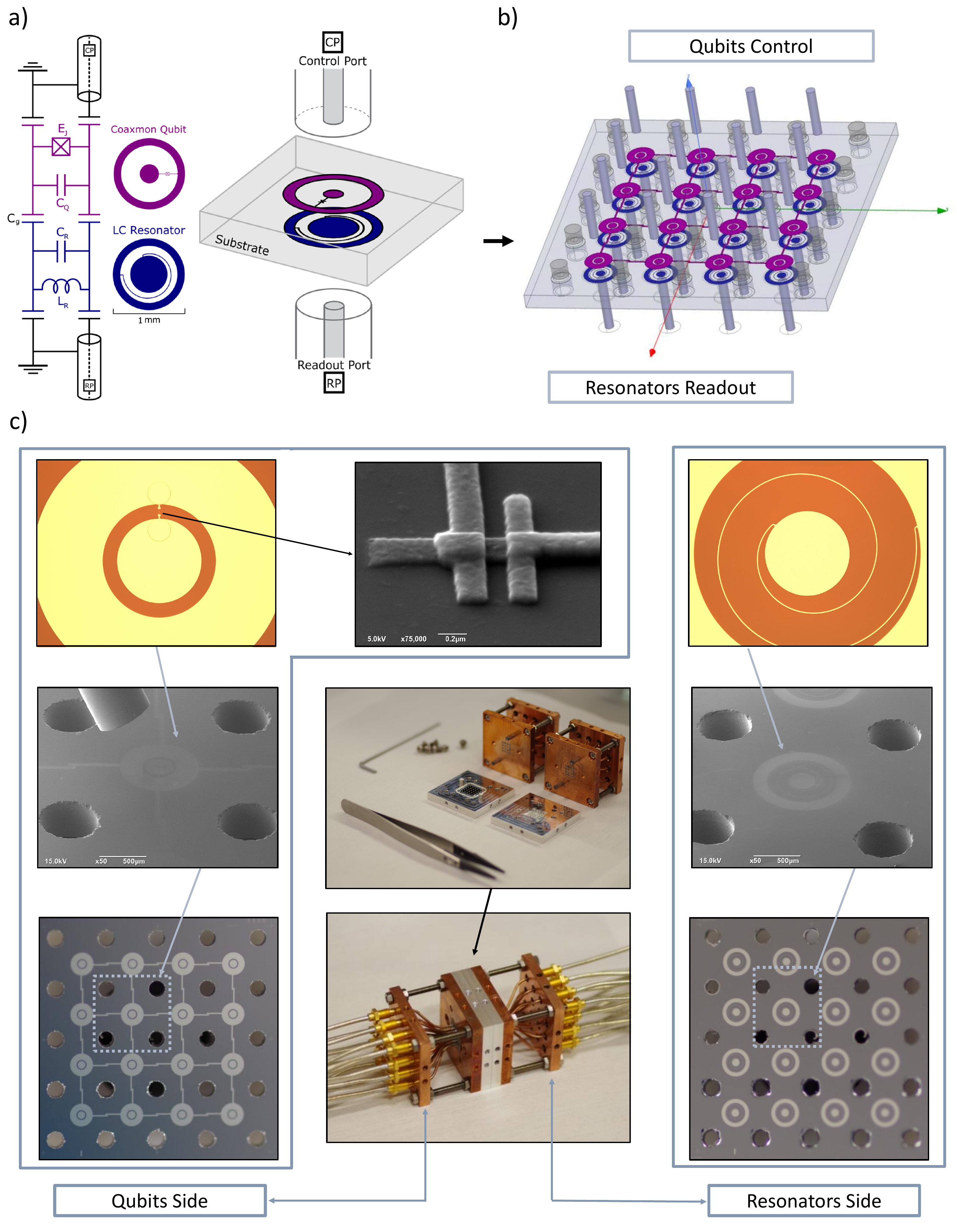}
      \caption{\textbf{Design Schematics and images of the fabricated device including enclosure packaging}. (a) 3D-integrated coaxial cQED architecture features a transmon qubit with a Josephson junction (JJ) and a readout resonator on the opposite side of the chip, enabling targeted couplings to off-chip coaxial ports for control and readout. (b) Extended design presented in this work consisting of 16 coupled qubits with capacitive arms, and metallic pillars for off-chip inductive shunting on the device enclosure. (c) Optical and SEM images of the qubits and resonators are shown, including key circuit elements and a single JJ. The device is fabricated using a double-sided process on a high-resistivity silicon wafer (see supplementary materials).
        }
  \label{F1}
\end{figure}

We demonstrate a scalable $4 \times 4$ square lattice of 16 fixed-frequency transmon qubits, implemented in a 3D-integrated cQED architecture \cite{rahamim2017double, spring2020modeling, spring2022high}. Each qubit is capacitively coupled through the substrate to a readout resonator positioned on the opposite side of the chip, featuring a tileable unit cell (see Fig.~\ref{F1} (a)). This approach enables individual qubit control and readout in a compact lattice architecture (see Fig.~\ref{F1} (b) and (c)). We implement off-chip inductive shunting on the device enclosure to mitigate box-mediated residual crosstalk originating from parasitic enclosure modes \cite{spring2020modeling, spring2022high}. Each qubit is capacitively coupled to its four nearest neighbors via lithographically patterned capacitive arms, facilitating interactions characterized by exchange energy rates $J_{i,j}$ in equ. \ref{eq.1} between each pair of qubits (see Fig.~\ref{F3} (a)). See supplementary materials for more details on the fabrication process and experimental setup. 

\subsection*{Basic Device Parameters}
\begin{figure}[H]
  \centering
   \includegraphics[width=1\textwidth]{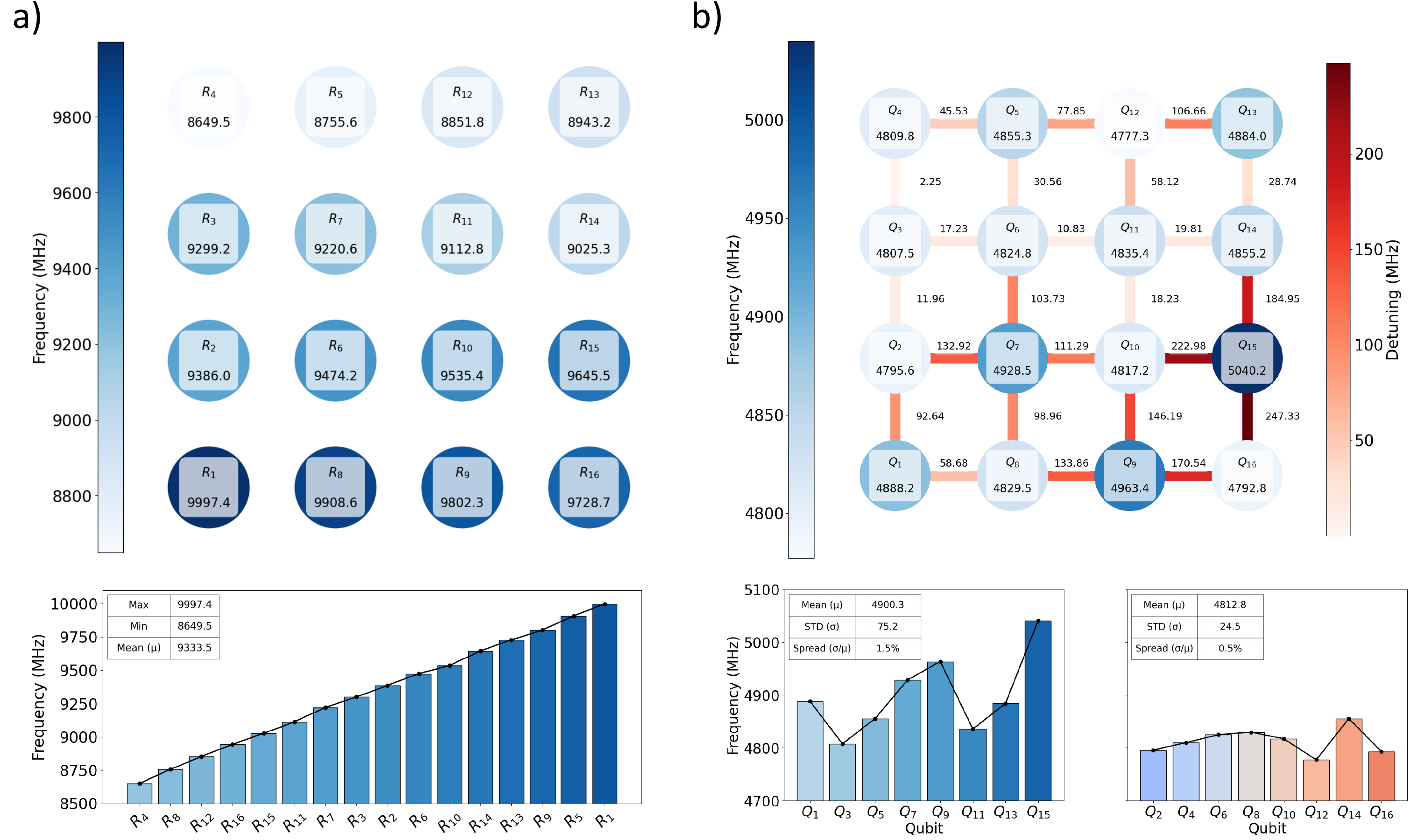}
    \caption{\textbf{Device parameters.} (a) Shows distinct resonator frequencies ranging from \textasciitilde{} 8.6 to 10 GHz, and (b) shows qubits with two-distinct, alternating, frequency pattern in a range between 4.8 and 4.9 GHz with most qubits operating in the straddling regime. This ensures that most qubits remain within a regime where their interactions can be effectively controlled for tuning up two-qubit interactions.
    }
  \label{F2}
\end{figure}

The resonators are designed to have distinct frequencies following a ladder design ranging from \textasciitilde{} 8.6 to 10.0 GHz, targeting well-separated readout frequencies (see Fig. \ref{F2}(a)). Each set of 8 qubits is designed with two-distinct, alternating, frequency pattern in a range between 4.8 and 4.9 GHz, and measured with very low frequency spreads of $0.5\%$ and $1.5\%$ MHz, receptively, for both targeted values (see Fig. \ref{F2}(b)). This frequency range allows operating in the straddling regime, where detunings of qubits remain smaller than their anharmonicities, necessary later for tuning up two-qubit interactions. We note that only two pairs of qubits, $Q_{15}$-$Q_{10}$ and $Q_{15}$-$Q_{16}$, are outside the straddling regime due to a higher frequency of $Q_{15}$. The average anharmonicity is $\langle \alpha \rangle_{16} = 196.4$ MHz across all qubits with a very low frequency spread of $0.6\%$ (see supplementary materials for more device parameters). The low frequency spreads were achieved without any further post-fabrication process on the junctions, such as junctions annealing \cite{hertzberg2021laser}, but only by fine tuning the junctions fabrication parameters.

\section*{Results and Discussion}

\begin{figure}[H]
  \centering
   \includegraphics[width=1\textwidth]{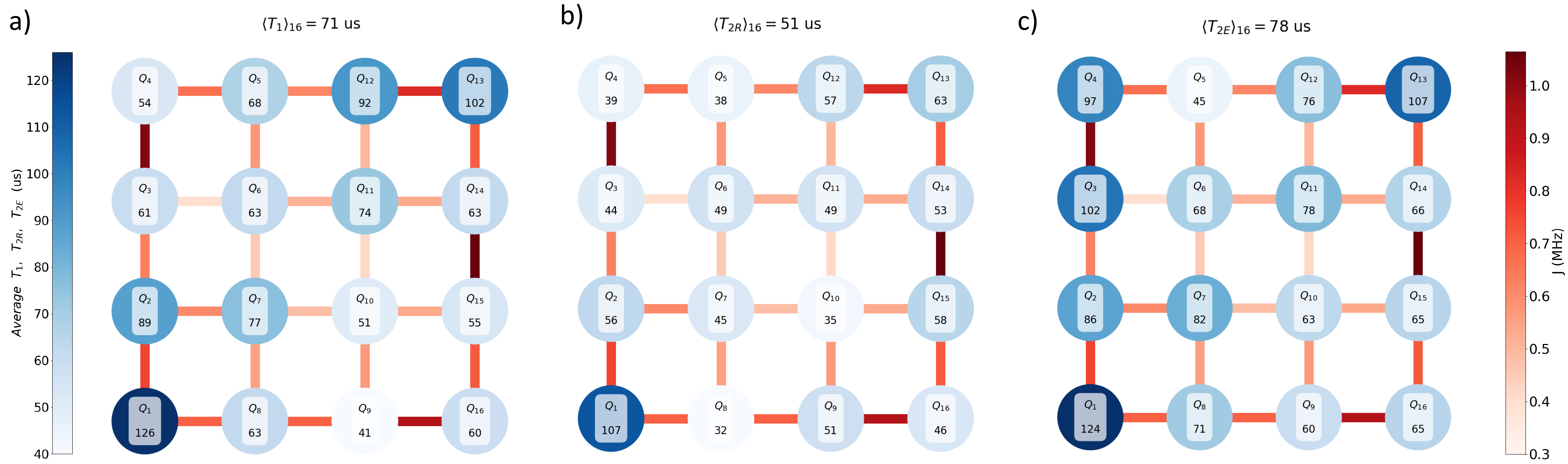}
  \caption{\textbf{Relaxation and coherence times}. (a) Energy relaxation times $\langle T_{1} \rangle_{400}$ averaged from measured traces following an exponential decay, fitted to $S(\Delta t) = a + b e^{-\Delta t/T_{1}}$. (b) Ramsey coherence times $\langle T_{2R} \rangle_{400}$ averaged from measured traces following a decaying oscillation, fitted to $S(\Delta t) = a + b \cos(2\pi \Delta f \Delta t + \phi) e^{-\Delta t/T_{2R}}$. (c) Spin-echo coherence times $\langle T_{2E} \rangle_{400}$ averaged from measured traces following an exponential decay, fitted to $S(\Delta t) = a + b e^{-\Delta t/T_{2E}}$. Each coupling $J_{i,j}$ across the lattice is shown with the colored bar representing the coupling strength between qubits.}
  \label{F3}
\end{figure}

Qubits relaxation and coherence times were measured repeatedly over 12 hours, resulting in a total of 400 measurements for each $T_{1}$, $T_{2R}$ and $T_{2E}$ that are then averaged for each qubit (see Fig. \ref{F3}). The average qubits relaxation times $T_1$ are shown in Fig.\ref{F3}(a), and dephasing times $T_{2R}$ and $T_{2E}$ were measured using Ramsey and Hahn echo sequences and shown in Fig. 3(b) and Fig. 3(c), respectively. Relaxation and coherence times averaged across the lattice are  $\langle T_{1} \rangle_{16} = 71 ~\pm ~5 ~\mu s$, $\langle T_{2R} \rangle_{16} = 51 ~\pm ~4 ~\mu s$ and $\langle T_{2E} \rangle_{16} = 78 ~\pm ~5 ~\mu s$, with weighted-standard deviations.

\subsubsection*{Crosstalk Characterization}
\begin{figure}[H]
  \centering
   \includegraphics[width=1\textwidth]{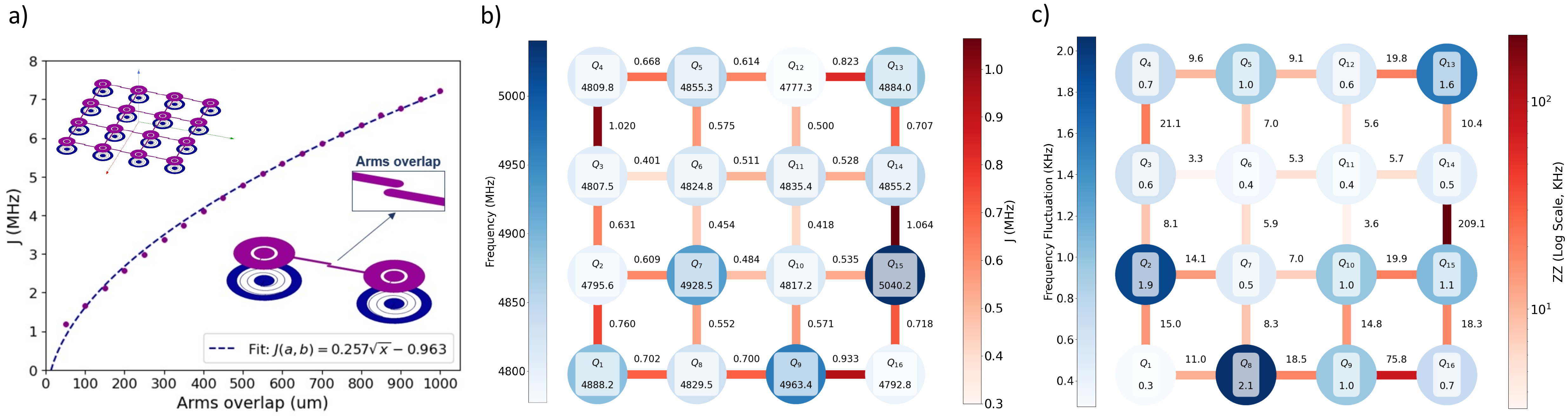}
  \caption{\textbf{Design of exchange coupling rate $J$ and nearest-neighbor ZZ and $J$ Measurements.}
        (a) Modeling of the coupling $J$ as a function of capacitive arms overlap obtained by combining a simple impedance formula \cite{solgun2019simple} with high-frequency structure simulator (HFSS) using terminal simulation. (b) The measured J values for nearest-neighbour pairs by performing Ramsey measurements between each pair of qubits and observing state-dependent ZZ frequency shifts in (c). The level of frequency fluctuations in each qubit is shown in (c) and measured by continuously performing $\sim$ 400 Ramsey experiments over \textasciitilde{} 12~hours and extracting the standard deviation in each qubit's frequency, achieving an overall average frequency fluctuations of 0.88 KHz across the device.}
  \label{F4}
\end{figure}

We present two methods to characterize direct coupling and crosstalk. One is based on ZZ measurements using equ. \ref{eq.3} and the second is based on direct measurements of anticrossing between different pairs of qubits using AC-Stark shift ~\cite{wallraff2007sideband}. Fig. \ref{F4} shows both the design and measurements of the coupling \(J\) between nearest-neighbor qubits. Fig. \ref{F4}(a) shows how a simple impedance-based model \cite{solgun2019simple} is used combined with HFSS simulations to model \(J\) as a function of the capacitive arm overlap. In Fig.~\ref{F4} (b) and (c), the measured ZZ shifts are used to calculate \(J\) values across the device using equ. \ref{eq.3}. The results also highlight each qubit’s frequency fluctuation, determined by \textasciitilde{} 400 repeated Ramsey measurements over \textasciitilde{} 12~hours, showing a very low frequency instability of about 0.88\,KHz as an average of all frequency fluctuations across the device (see Fig.~\ref{F4} (c)).

In particular, we observe low ZZ shifts across all qubit pairs except for two notable outliers occur for $Q_{9}$-$Q_{16}$ (75.8\,KHz) and $Q_{14}$-$Q_{15}$ (209.1\,KHz), which lie near higher transitions and near the edge of the straddling regime. The estimated couplings have maximum and minimum values at 1.064\,MHz and 0.401\,MHz, respectively. The mean \(\mu\) of 0.623\,MHz and standard deviation \(\sigma\) of 0.173\,MHz lead to a relative spread \(\sigma/\mu\) of about 0.269 in different values of the coupling \(J\) across the lattice.

\begin{figure}[H]
  \centering
   \includegraphics[width=1\textwidth]{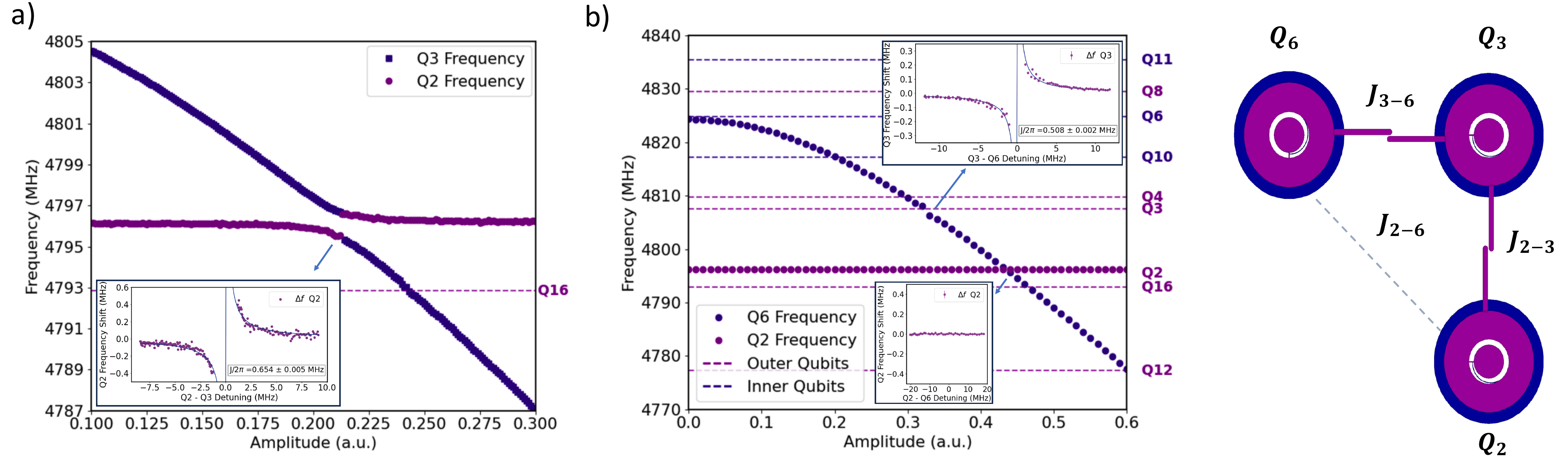} 
  \caption{\textbf{Nearest-neighbor coupling $J_{i,j}$ and long-range crosstalk \(\widetilde{J}_{i,j}\).} 
    Both direct (a) and indirect (b) couplings are measured using AC Stark Ramsey between each pair of qubit in the three-qubit example. Each qubit is stark shifted with an additional AC tone until it is in resonance with the other qubit, and the observed frequencies are fitted to a simple anticrossing model given by \( \Delta f_{AC} = A \frac{J_{i,j}^2}{B ~ \Delta_{i,j}} + C \), where \( J_{i,j} \) is the exchange coupling between qubits, \( \Delta \) is the detuning between the qubits, and \( A \), \( B \), and \( C \) are fitting parameters. 
    }
  \label{F5}
\end{figure}

\begin{figure}[H]
  \centering
   \includegraphics[width=1\textwidth]{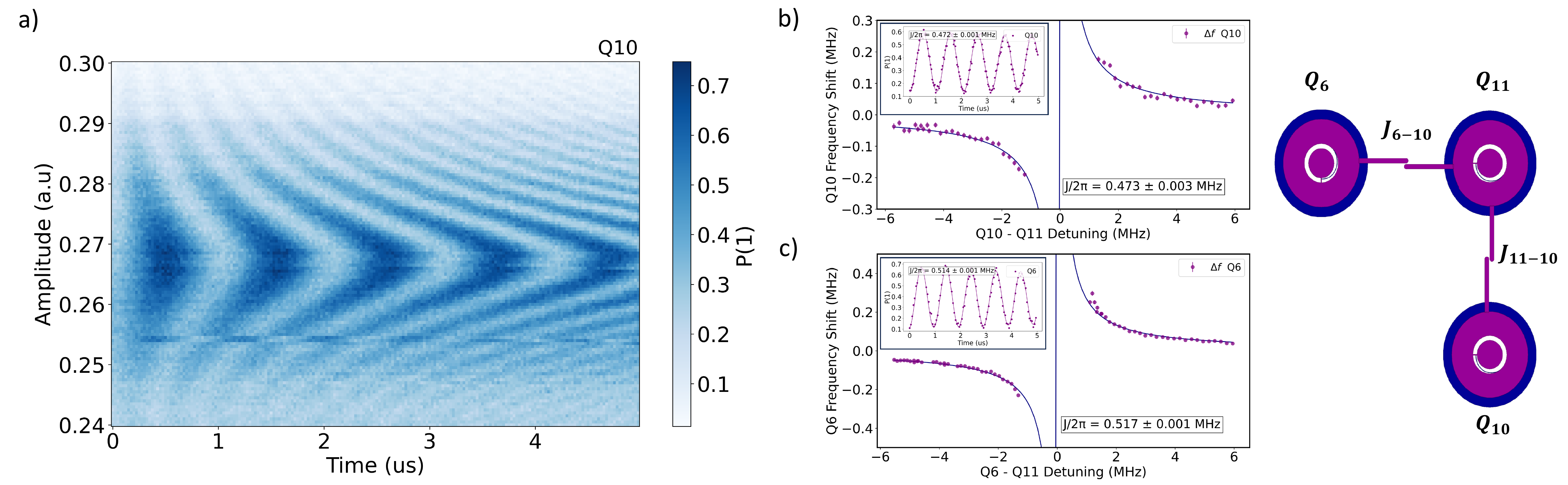} 
  \caption{\textbf{Nearest-neighbor coupling $J_{i,j}$.} 
    Set of three-qubit in the lattice is considered to verify the measured coupling values through observing both dynamics of swap population (a) and anticrossing from AC Stark-shift Ramsey (b). In (a) one qubit is initially excited and then Stark-shifted until a coherent swap of population is observed once the qubit is in resonance with its nearest-neighbor qubit. In (b) and (c), the same exchange rates are extracted independently using AC Stark Ramsey, where one qubit is Stark shifted until the qubit is in resonance with its neighbor and the frequency shift is fitted to extract the exchange energy rate. The inset figures in (b) and (c) show a coherent swap of populations once both qubits are in resonance at the same rate estimated from AC Stark-shift Ramsey. 
    }
  \label{F5_P1}
\end{figure}

Next, direct measurements of anticrossing between different pairs of qubits using AC-Stark shift are shown in Fig. \ref{F5}. This example illustrates both direct and indirect interactions in a simplified three-qubit setting in the lattice. In Fig. \ref{F5}(a), one qubit is Stark-shifted until it is in resonance with its nearest neighbor, forming an avoided-crossing from which the exchange coupling \(J_{i,j}\) is directly extracted by fitting to a simple model. Table~\ref{tab:swap_nn_qubits} compares the resulting \( J_{\mathrm{SWAP}} \) values for nearest neighbors obtained by performing AC-Stark shift Ramsey experiments, with those extracted from measuring static ZZ shifts, \( J_{ZZ} \). We confirm these measurements by performing AC-Stark shift on another set of three qubit in lattice, and observe a swap of population between pairs of nearest-neighbor qubits (see Fig. \ref{F5_P1}(a)), and compare the exchange energy rates with those extracted from AC-Stark shift Ramsey measurements on the same pairs of qubits.

\begin{table}[htbp]
    \centering
    \caption{Characterization of the nearest-neighbor ${J}_{i,j}$ couplings from anticrossings using AC-Stark shift Ramsey compared to estimated coupling values from measuring static ZZ shifts in Fig. \ref{F4}(c).}
    \resizebox{0.55 \textwidth}{!}{ 
        \begin{tabular}{|c|c|c|c|c|c|c|c|c|c|}
            \hline
            Qubits & \( w_{q_A}/2\pi \) & \( w_{q_B}/2\pi \) & \( \Delta_{AB} \) & \( ZZ_{static} \) & \( J_{ZZ} \)  & \( J_{SWAP} \) \\

            pair & MHz & MHz & MHz & MHz & MHz & MHz  \\
            \hline
            \( Q_{2}-Q_{3} \)   & 4795.6 & 4807.5 & 12.0 & 0.0081  & 0.631  & 0.654  \\ 
            \( Q_{3}-Q_{6} \)   & 4807.5 & 4824.8 & 17.2 & 0.0033  & 0.401  & 0.508  \\ 
            \( Q_{6}-Q_{11} \)  & 4824.8 & 4835.4 & 10.7 & 0.0053  & 0.511  & 0.517  \\ 
            \( Q_{10}-Q_{11} \) & 4835.4 & 4817.2 & 18.2 & 0.0036  & 0.418  & 0.473  \\ 
            \( Q_{14}-Q_{11} \) & 4835.4 & 4855.2 & 19.8 & 0.0057  & 0.528  & 0.536  \\ 
            \hline
        \end{tabular}
    }
    \label{tab:swap_nn_qubits}
\end{table}


In addition, in Fig. \ref{F5}(b), the same measurements are performed on diagonal pairs of qubits that are not directly connected by a capacitive arm, in which any measurable residual coupling represents crosstalk in this case. This long-range interaction arises from higher-order virtual processes or enclosure-mediated parasitic effects. In Fig. ~\ref{F5}(b), we observe no long-range couplings across the lattice as one qubit is Stark-shifted across multiple adjacent qubits in frequency and space. The same observation is seen on multiple other cases summarized in Table~\ref{tab:swap_non_nn_qubits}. All measured values represents the frequency fluctuations from Ramsey as shown in the inset measurements in Fig. \ref{F5}(b), which all remain significantly lower than the measured couplings for nearest neighbors, indicating that parasitic crosstalk is very small in the device. This supports the observation that the dominant couplings in the device are the deliberately engineered nearest-neighbor interactions and that spurious or long-range crosstalk can be kept well near the intrinsic frequency fluctuation levels of each qubit (see Fig. ~\ref{F4}(c)).

\begin{table}[htbp]
    \centering
    \caption{Characterization of the non-nearest-neighbor $\widetilde{J}_{i,j}$ crosstalk from anticrossings using AC-Stark shift Ramsey.}
    \resizebox{0.45\textwidth}{!}{ 
        \begin{tabular}{|c|c|c|c|c|c|c|c|c|}
            \hline
            Qubits & \( w_{q_A}/2\pi \) & \( w_{q_B}/2\pi \) & \( \Delta_{AB} \) & \( \text{Std-Dev} \) \\
            pair & MHz & MHz & MHz & MHz  \\
            \hline
            \( Q_{6}-Q_{2} \)   & 4795.6 & 4824.8 & 29.2 &  0.00507 \\ 
            \( Q_{6}-Q_{10} \)  & 4824.8 & 4817.2 & 7.6 &  0.01200 \\ 
            \( Q_{5}-Q_{11} \)  & 4855.1 & 4835.4 & 19.7 &  0.03927 \\ 
            \( Q_{8}-Q_{10} \)  & 4829.5 & 4817.2 & 12.3 &  0.00807 \\ 
            \( Q_{10}-Q_{16} \) & 4817.2 & 4792.8 & 24.3 &  0.00662 \\ 
            \hline
        \end{tabular}
    }
    \label{tab:swap_non_nn_qubits}
\end{table}


\subsubsection*{Single-Qubit Gate Errors}

\begin{figure}[H]
  \centering
   \includegraphics[width=1\textwidth]{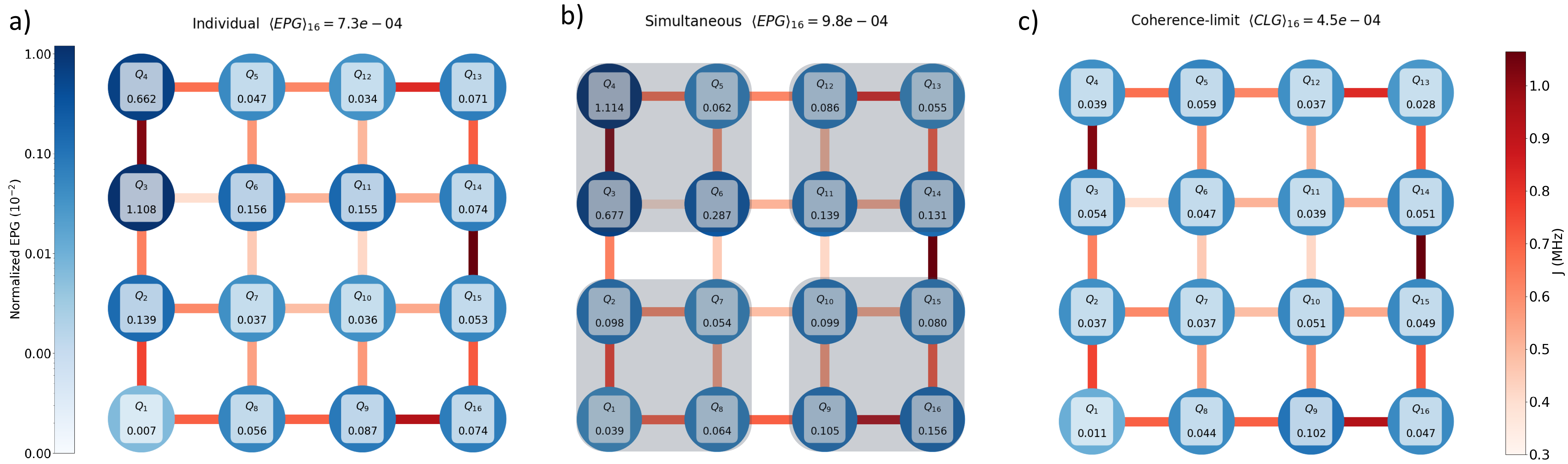}
  \caption{\textbf{Single-qubit gate errors}. (a) Individual and (b) simultaneous error-per physical gate errors (EPG) with median values across the device compared to coherence-limit physical gate errors (CLG) in (c). Measurements of EPG is done by randomized benchmarking on $XY$ Clifford decomposition for both individual and simultaneous measurements across the lattice. Each coupling $J_{i,j}$ across the lattice is shown with the colored bar representing the coupling strength between qubits.} 

  \label{F6}
\end{figure}

Single-qubit gate errors across the device are shown in Fig. \ref{F6}, evaluated through randomized benchmarking (RB) ~\cite{chow2009randomized, gambetta2012characterization}. Fig.\ref{F6}(a) shows the error-per-physical-gate (EPG) values for individual qubits, while Fig. \ref{F6}(b) shows simultaneous measurements on four-qubit sets across the lattice. RB experiments were conducted using an \( XY \)-Clifford decomposition for both individual and simultaneous gate errors. The resulting error rates were obtained by applying a combination of 60~ns duration (consisting of 50~ns Blackman envelope with 10~ns buffer) of $I$, $X_{\pi/2 ,\pi}$ physical gates, combined with derivative removal gate (DRAG) pulse shaping \cite{motzoi2009simple} and virtual Z gates \cite{mckay2017efficient}. Single-shot readout was performed during all RB experiments with readout time of 8~us. 

For both individual and simultaneous RB, each experiment was performed with 16 different Clifford sequences with total sequence length of 1000 gates and with each sequence repeated for \( N = 10 \) distinct Clifford gates. Each of the 16 × 10 experiments was performed on every qubits for both individual and simultaneous RB (see supplementary materials). The EPG values obtained from these experiments are summarized Fig. \ref{F6}(a) and (b). To compare the measured gate errors with coherent limits, the coherence-limited EPG (denoted as CLG) is shown in Fig. \ref{F6}(c) and calculated using the expression \( CLG = (3 - \exp(-t_g/T_1) - 2\exp(-t_g/T_{2E}))/6 \) ~\cite{solgun2019simple}, where \( t_g \) is the total duration of each physical gate (60~ ns), and \( T_1 \) and \( T_{2E} \) represent the relaxation and echo coherence times, respectively. This theoretical limit provides a benchmark to evaluate the fidelity of the single-qubit gates in relation to the intrinsic coherent errors. We observe very low median gate errors across the lattice and comparable EPGs on both individual and simultaneous RB experiments except for the pair of qubits $Q_{3}$-$Q_{4}$ in which both qubits have higher errors due to the fact that this pair has very low detuning of 2.25 MHz (see Fig. \ref{F2}(b)). Importantly, we observe comparable simultaneous (to individual) single-qubit gate errors across the lattice, despite the presence of always-on qubit-qubit coupling. This shows that correlated errors arising from residual crosstalk remain suppressed in the device, indicating minimal error propagation between qubits during simultaneous gates operations.


\subsubsection*{Two-Qubit Interactions and CZ Gates}

\begin{figure}[H]
  \centering
  \includegraphics[width=0.9\textwidth]{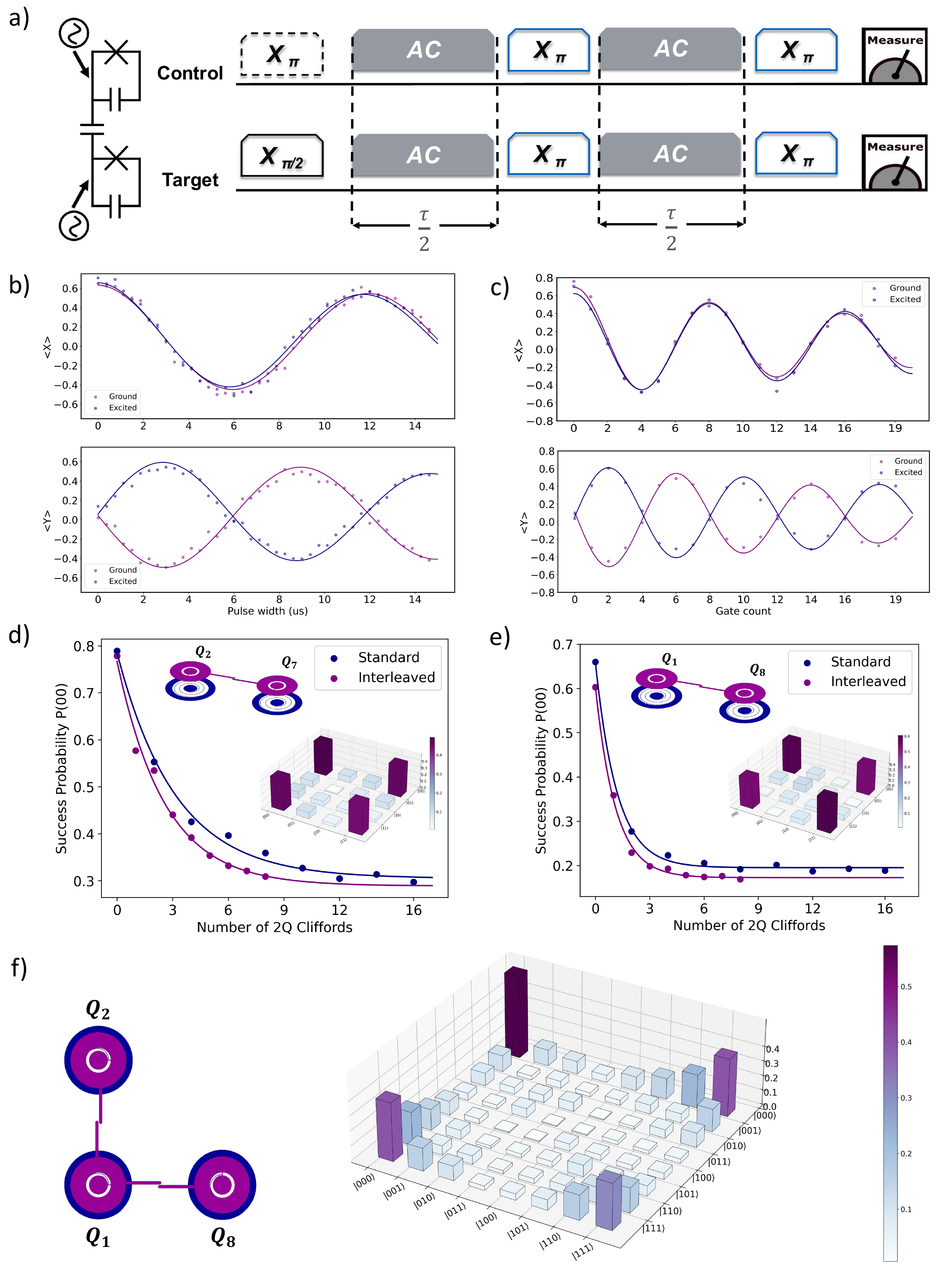}
  \caption{\textbf{Two-qubit interactions and CZ gates Calibration}. 
        (a) Pulse sequence implemented to tune up two-qubit ZZ interactions based on the siZZle technique. Two-qubit ZZ-induced phase accumulation on target qubit for calibrating a CZ gate on $Q_{2}$ (control) and $Q_{7}$ (target) as function of AC pulse width is shown in (b) and as function of gate count is shown in (c) after setting up an optimal gate duration. Two-qubit randomized benchmarking and Bell state preparation between $Q_{2}$ and $Q_{7}$ are shown in (d), and between $Q_{1}$ and $Q_{8}$ are shown in (e). (f) Shows preparation of GHZ state between $Q_{1}$, $Q_{2}$ and $Q_{8}$, demonstrating a multi-qubit entanglement across the lattice.}
  \label{F9}
\end{figure}

We implement entangling operations between fixed-frequency transmon qubits in the lattice by using the Stark-induced ZZ by level excursions (siZZle) technique \cite{PhysRevLettMitchell, PhysRevLettWei} to boost static ZZ coupling. Here, we use two additional off-resonant drives to induce parametrized shifts in the energy levels of a two-transmons system as shown in Fig. ~\ref{F9}(a). This approach modifies the native ZZ interaction and can be used to tune up a controlled-Z (CZ) gate by driving each transmon with a detuned microwave tone. The two simultaneous off-resonant drives on the two qubits shift the energy levels of the system and, through the capacitive coupling, modify the effective ZZ rate between the qubits. The modified ZZ rate $\tilde{\nu}_{ZZ}$ can be approximated as \cite{wei2021quantum}:

\begin{equation}
\tilde{\nu}_{ZZ} = \nu_{ZZ,s} + \frac{2J \alpha_0 \alpha_1 \Omega_0 \Omega_1 \cos(\phi_0 - \phi_1)}{\Delta_{0,d} \Delta_{1,d} (\Delta_{0,d} + \alpha_0)(\Delta_{1,d} + \alpha_1)},
\label{eq:ZZ_sizzle}
\end{equation}

where $\nu_{ZZ,s} = \zeta$ denotes the static ZZ interaction rate in equ.\ref{eq.3}, $\alpha_0$ and $\alpha_1$ are the anharmonicities, $J$ is the fixed capacitive coupling strength, and $\Omega_0$, $\Omega_1$, $\Delta_{0,d}$, $\Delta_{1,d}$, $\phi_0$ and $\phi_1$ denote the drive amplitudes, detunings to each qubit, and relative phases of drives, respectively. In principle, the effective $\tilde{\nu}_{ZZ}$ rate can be boosted or canceled depending on the sign of the additional driving term. The calibration of a CZ gate based on the siZZle interaction requires optimizing the drive parameters such that the total ZZ-induced phase accumulation during the gate operation equals $\pi/4$. We set the two drive amplitudes to be relatively equal to $\Omega_{control} = r ~\Omega_{target}$, for maximum $\tilde{\nu}_{ZZ}$ while observing a clean interaction, and $r$ here is the ratio between the amplitudes of the single-qubit $X_{\pi}$ pulses for the control to the target qubits. We have found this relation takes into account the asymmetries between the two drive amplitudes, introduced by cabling or room-temperature electronics. The relative phase between the two drives was found to be near-optimal in our setup and was set to be $\Delta = \phi_0 - \phi_1 = 0$ during CZ gate calibration. See supplementary materials for more details about tuning up two-qubit interactions and calibrating CZ gates. 

Pulse sequence for siZZle gate calibration is shown in Fig. ~\ref{F9}(a). The siZZle gate is implemented using two off-resonant Stark drives, with interleaved and final single-qubit $\pi$ pulses to cancel unwanted single-qubit phase accumulation. A dashed line on the the single-qubit pulse on the control in Fig. ~\ref{F9}(a) denotes that the experiment is run both with and without exciting the control. In Fig. ~\ref{F9}(b), pulse width Hamiltonian tomography on the target qubit is shown to extract the ZZ interaction. The duration of the Stark pulse here is swept while monitoring the phase evolution of the target qubit. In Fig. ~\ref{F9}(c), repeated gate Hamiltonian tomography on the target qubit is shown. This experiment now uses fixed-duration two-qubit pulses and repetition blocks to more precisely calibrate the accumulated ZZ phase, which can then be used to implement an entangling gate. States of the control qubit $Q_{2}$ during ZZ-induced phase accumulation on target qubit $Q_{7}$ (see in Fig. ~\ref{F9}(b)) are shown in Fig. ~\ref{F9b} in supplementary materials.

We verified the calibrated CZ gate by interleaving it into a two-qubit randomized benchmarking (RB) sequence \cite{cao2023emulating, cao2023implementation}, performed on multiple pairs across the lattice as shown in Fig. ~\ref{F9}. The single-qubit gates here were optimally calibrated with 20~ns duration (consisting of 16~ns Blackman envelope with 4~ns buffer) of $X_{\pi/2}$ physical gates, combined with derivative removal gate (DRAG) pulse shaping \cite{motzoi2009simple} and virtual Z gates \cite{mckay2017efficient}. Single-shot readout was also optimized during all two-qubit experiments with a readout time of 3~us. The gate calibration consists of oberving ZZ-induced phase accumulation during the gate operation on the target qubits as shown in Fig. ~\ref{F9}(b) and (c). The performance of the CZ gates was further complemented by the direct preparation of entangled Bell states between two qubits (in Fig. ~\ref{F9}(d) and (e)), and preparation of GHZ state between three qubits across the lattice as shown in Fig. ~\ref{F9}(f).

For the first pair in Fig. ~\ref{F9}(d) consisting of $Q_{2}$ (control) and $Q_{7}$ (target), we achieve CZ gate fidelity of $95.15 \pm 1.76~\%$ for a total gate time of $\tau_{g} = 3.266 ~\mu s$, resulting in an average Bell state fidelity of $93.56\%$ measured by two-qubit state tomography. For the second pair in Fig. ~\ref{F9}(e) consisting of $Q_{1}$ (control) and $Q_{8}$ (target), we achieve CZ gate fidelity of $96.44 \pm 1.78~\%$ for a total gate time of $\tau_{g} = 2.623 ~\mu s$, resulting in an average Bell state fidelity of $90.0\%$ measured by two-qubit state tomography. Finally, the prepared GHZ state between $Q_{1}$, $Q_{2}$ and $Q_{8}$ in Fig. ~\ref{F9}(f) has an average GHZ state fidelity of $83.88\%$ measured by three-qubit state tomography. 

\section*{Conclusion}
We have demonstrated a scalable 4×4 square lattice of 16 fixed-frequency transmon qubits with nearest-neighbor capacitive coupling, implemented in a tileable, 3D-integrated cQED architecture with off-chip inductive shunting. The device achieves well-targeted qubit frequencies with very low spreads across two distinct frequency groups. We characterized coupling and crosstalk using both static ZZ shifts and direct anticrossing measurements, confirming that inter-qubit couplings remain localized, with negligible long-range parasitic interactions. Simultaneous randomized benchmarking shows low single-qubit gate errors and comparable to individual gate errors, with median error rates approaching coherence-limited errors. These results validate our design approach and present a practical architecture for scaling superconducting quantum circuits with low crosstalk and high qubit connectivity.


\section*{Acknowledgment}
This work has received funding from the United Kingdom Engineering and Physical Sciences Research Council (EPSRC) under Grants No. EP/N015118/1, No. EP/T001062/1 and No. EP/W024772/1. M.B. acknowledges support from EPSRC QT Fellowship under Grant No. EP/W027992/1. S.C. acknowledges support from Schmidt Science. We would like to acknowledge the Superfab Nanofabrication facility at Royal Holloway, University of London, where part of device fabrication was performed.






\section*{Supplementary Materials}
Superconducting quantum circuits are commonly fabricated using thin-films of aluminum, niobium, or titanium alloys on silicon or sapphire substrates \cite{In13}. The most critical components of these circuits are Josephson junctions, which when shunted with large capacitors, can form the widely used type of superconducting qubits known as the transmon \cite{CQ1}. The fabrication process of these junctions can vary based on the junction size controlled by electron-beam lithography exposure dose, oxidation parameters, and metal evaporation pressure. In our study, we employ a double-sided fabrication on a double-side 3-inch polished intrinsic silicon wafers, involving multiple lithography steps, thin-film depositions, and protective resists to ensure high-quality surfaces and interfaces on both sides during the fabrication process. The qubits and resonators are fabricated on opposite sides of the silicon substrate and capacitively coupled through the bulk substrate. The coupling strength is primarily determined by the substrate thickness and the geometry of the capacitive pads of both the qubit and the resonator \cite{rahamim2017double}.

\subsection*{Fabrication Process}
\label{subsec:Fabrication}
The detailed steps of the fabrication process are described \cite{alghadeer2025characterization, peterer2016experiments, cao2023implementation}, with relevant design parameters given in Table ~\ref{Table.1}. Spin-coating a protective photoresist layer on the backside is critical in double-sided fabrication process to protect the wafer and prevent additional contamination. The resonators side is patterned first while the qubit side is covered with photoresist, followed by cleaning the photoresist and spin-coating another protective photoresist layer on the resonators side and patterning the qubits side.

\begin{table}[H]
  \centering
  \caption{Geometric design parameters of relevant parts of the device shown in Fig.~\ref{F1}.}
  \label{Table.1}
  \begin{tabular}{Sl Sl Sl}
    \hline
    Geometric parameter          &   Symbol    & Standard value                      \\
    \hline
    Spiral line width            & ~~~~~ $s$       &   5 ~~~~~~~~~~~~~~~~~~~~   µm       \\
    Al thin film thickness       & ~~~~~ $t$       &   100   ~~~~~~~~~~~~~~~~   nm       \\
    JJ thin film thickness   & ~~~~~ $d$       &   (27-30)  +  70      ~~   nm       \\
    Si substrate thickness       & ~~~~~ $h$       &   500   ~~~~~~~~~~~~~~~~   µm       \\
    \hline
  \end{tabular}
\end{table}

\subsubsection*{Wafer Cleaning}

The fabrication process starts with cleaning a high-resistivity (\(>\,10\,\mathrm{K\Omega\cdot cm}\)) intrinsic silicon wafer using a 10:1 buffered oxide etch (BOE) solution of hydrofluoric acid and ammonium fluoride to remove native oxides and contaminants. After etching, the wafers are thoroughly rinsed with ultrapure deionized water, dried with nitrogen gas, and promptly transferred (within 5~min) to minimize re-oxidation before thin-film deposition.

\subsubsection*{Aluminum Thin-Film Deposition}

Immediately after water cleaning, the wafer is immediately loaded in Plassys MEB550S2 at ultra-high vacuum (UHV) and is baked up to $200\,^\circ\mathrm{C}$ for 10~min. After which a layer of 100 nm of aluminum is deposited at rate of 1 nm/s on the substrate through UHV electron-beam evaporation under controlled temperature and low-pressure conditions, with a base pressure down to $10^{-9}$~mbar and an evaporation pressure of around $10^{-8}$~mbar, ensuring high purity and uniformity of the thin film. The deposition rate and substrate temperature are carefully controlled to ensure smooth thin-film growth for optimal grain structure.

\subsubsection*{Photolithography and Micro-scale Circuit Elements}

A positive photoresist AZ~1514~H is spin-coated onto the wafer and then exposed to ultraviolet light through a chrome photomask that defines the desired circuit patterns. After development with AZ 726 MIF developer solution, the exposed areas of aluminum are revealed for etching. The aluminum is then selectively etched away using a wet etching process to define the circuit elements. An aluminum etchant Alfa Aesar 44581 solution and water are used to achieve anisotropic etching with optimal selectivity to minimize remaining aluminum defects. This step creates the micro-scale features of the circuit, including capacitors, inductors, and coupling interconnects. Immediately after the etching process, residual resist is removed using DMSO.

\subsubsection*{Electron-Beam Lithography and Nano-scale Josephson Junctions}

For the nano-scale features, high-resolution electron-beam lithography (EBL) is used to define the Josephson junctions. The junctions are fabricated using the Dolan bridge technique \cite{dolan1988very}, which involves double-angle evaporation of aluminum to form the tunnel barriers, followed by careful removal of excess aluminum through a lift-off process. A bilayer resist structure is employed, consisting of a copolymer (MA/MMA) and a polymethyl methacrylate (PMMA) layer, to create an undercut profile necessary for the shadow evaporation process. After spin-coating the resist, EBL is carried out in a JEOL system at 100~keV, using aperture Ap4 size 2~nA - 60~$\mathrm{\mu m^2}$ for small features and Ap8 size 100~nA - 300~$\mathrm{\mu m^2}$ for large features, with doses typically around 1500~$\mathrm{\mu C/cm^2}$. Following the exposure, the critical features are then developed using a mixture of IPA/MIBK mixture in a 3:1 ratio.

After EBL patterning, the wafer is loaded into the Plassys MEB550S2. Prior to deposition, an argon (Ar) ion milling is performed for 1~min (voltage 400~V, acceleration voltage 90~V/s, current 15~mA) to remove any residual contaminants and native oxides from the metal and substrate surfaces, ensuring a clean interface for the subsequent aluminum deposition. The first layer of junction is then deposited at an angle of $60^\circ$ from normal incidence, depositing 60~nm of Al at a rate of 0.5~nm/s. Due to the deposition angle, the effective thickness of the deposited film is approximately 27-30~nm. Following the first deposition, an \textit{in situ} controlled static oxidation inside Plassys is performed, typically for 5-10~min at an oxygen pressure of 5-10~mbar, depending on the target junction resistance. This controlled oxidation forms the thin insulating barrier of aluminum oxide essential for the tunnel junction. After pumping back down to UHV conditions, the second layer of aluminum is deposited at normal incidence ($0^\circ$), depositing 70~nm of Al at a rate of 0.5~nm/s, effectively completing the Josephson junction structure. Precise control over the oxidation parameters, such as oxygen pressure and exposure time, is critical to achieve the desired tunnel barrier properties and, consequently, the critical current of the junction \cite{dolan1988very}. Following evaporation, a lift-off process is carried out in a DMSO solution at $80\,^\circ\mathrm{C}$ for around 2~hrs and immediately followed by thoroughly rinsing with ultrapure deionized water and drying using nitrogen gas. 

\subsubsection*{Post-Fabrication Milling, Dicing and Packaging}

The wafer is next spin-coated with a protective layer of AZ~1514~H photoresist on both sides to protect the surfaces during milling and dicing. For milling, a central aperture of 500\,\textmu m diameter is drilled with a diamond micro-grinding tool using a Loxham Precision \textmu6 micro-machining system. These micromachining steps require careful handling to avoid introducing contamination or mechanical damage on the wafer, which could lead to additional defects. After milling, the wafer is diced into individual square dies of approximately 10\,\text mm side length using a Disco DAD3430 dicing saw. A diced chip is then mounted into a sample holder and prepared to be installed into a cryostat for microwave measurements.

\subsection*{Experimental Setup}
\begin{figure}[!ht]
        \centering
        \includegraphics[width=0.9\linewidth]{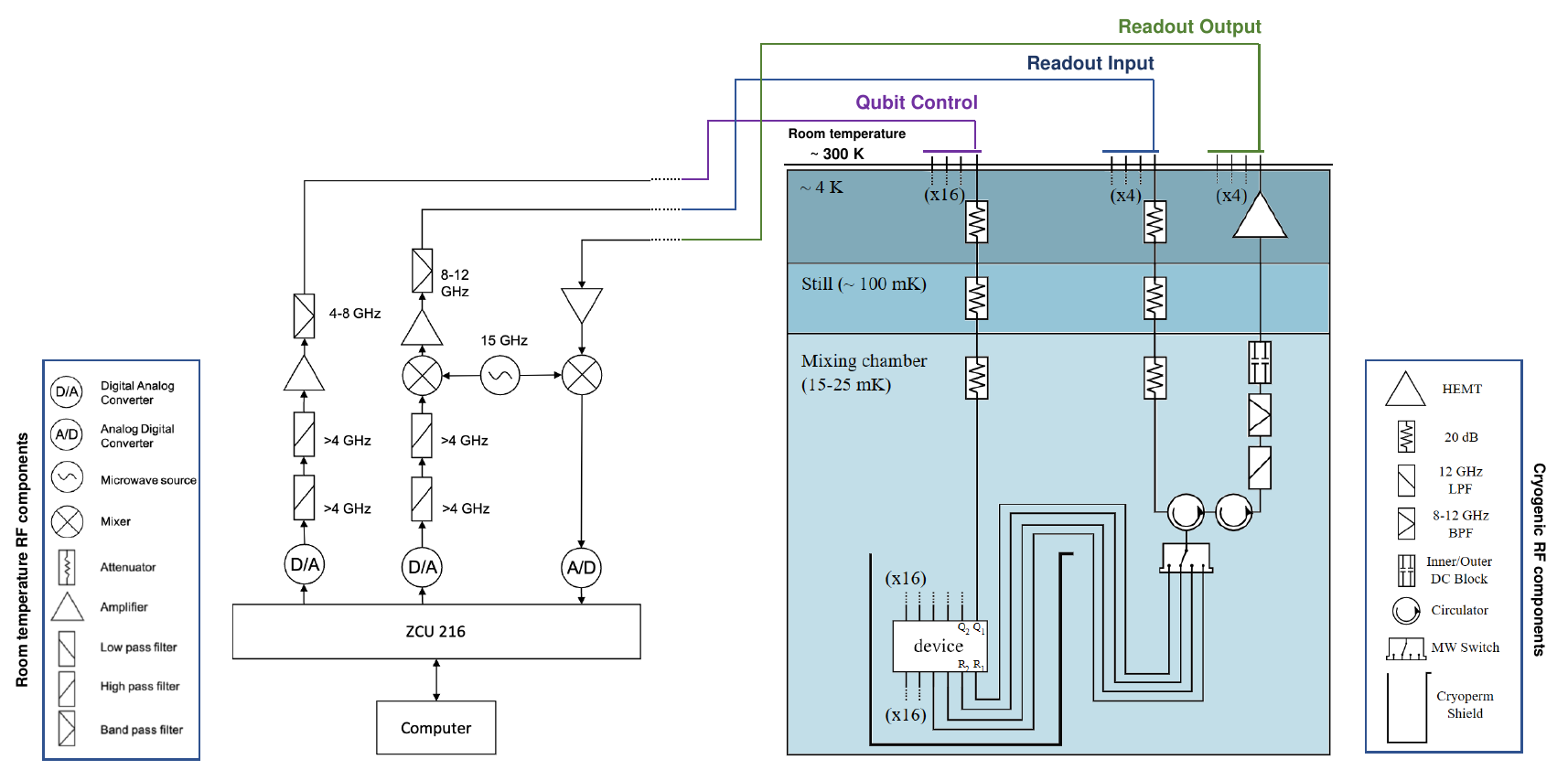}
        \caption{\textbf{Schematic diagrams outlining the experimental setup and dilution refrigerator with wiring and cryogenic components}. (a) Basic measurements for qubits and resonators are done using a Vector Network Analyzer (VNA) using the illustrated connections mainly for measuring resonance signals of resonators and qubits. Setup for a more precise pulsed qubit control generated by a set of FPGA boards with room temperature electronics and components illustrated.}
        \label{Experiemntal_Setup_XL}
    \end{figure}

The experimental setup operates at a base temperature of $\sim$ \( 15 \, \text{mK} \) using a \( ^3\text{He}/^4\text{He} \) dilution refrigerator. The control and readout of qubits are facilitated by the wiring configurations shown in Figure \ref{Experiemntal_Setup_XL}. For qubit control, microwave pulses are synthesized directly using QubiC system \cite{xu2021qubic, huang2023qubic}. The pulses are carefully shaped for single and two-qubit gates to align at the desired frequencies. For readout, reflected signals from the resonators go through amplification and down-conversion and are then captured by Analog-to-Digital Converters (ADCs) connected to the FPGA for measurements and further data analysis. The system is equipped with cryogenic attenuators and low-pass filters along the input lines to minimize thermal noise and spurious signals reaching the qubits. Output lines are similarly configured with isolators and cryogenic HEMT amplifiers to preserve signal integrity and maintain high signal-to-noise ratio throughout the measurement chain.

\newpage
\subsection*{Additional Device Parameters}
\begin{figure}[!ht]
        \centering
        \includegraphics[width=0.85\linewidth]{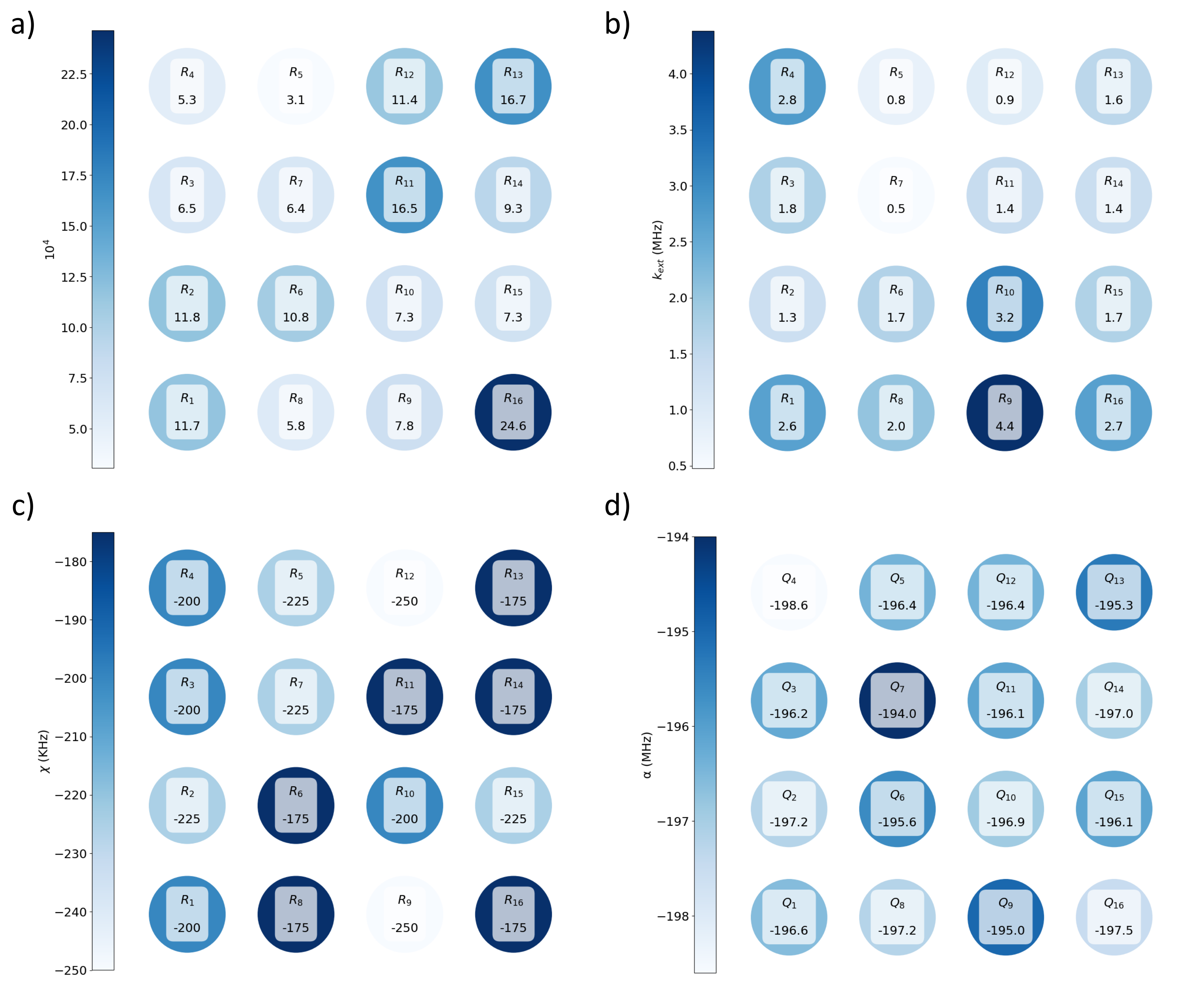}
        \caption{\textbf{Additional Device Parameters}. (a) $Q_{\text{int}}$ is internal Q factor of each resonator, (b) $\kappa_{\text{ext}}$ is external decay rate of each resonator, (c) $\alpha$ is the anharmonicity  of each qubit, and (d) $\chi$ is the dispersive shift of each qubit.}
        \label{FA_SupMat}
    \end{figure}

The following are the full datasets used to produce the results presented in the main text and more device parameters. Additional device parameters are visualized in Fig. \ref{FA_SupMat} and given in Table. \ref{basic_device_parameters}.

\begin{table}[!ht]
    \centering
    \caption{Basic Device Parameters and microwave characterization. \( w_{r}/2\pi \) and \( w_{q}/2\pi \) are frequencies of the readout resonator and qubit, respectively. \( Q_{i} \) is the internal quality factor of the resonator, and \( \kappa_{\text{ext}} \) is the external coupling rate. \( \chi \) is the qubit-resonator dispersive shift, and \( \alpha \) is the qubit anharmonicity. Qubits relaxation and coherence times $T_{1}$, $T_{2R}$ and $T_{2E}$ are averaged over 400 repeated measurements.}
    \resizebox{0.8\textwidth}{!}{ 
        \begin{tabular}{|c|c|c|c|c|c|c|c|c|c|}
            \hline
            Parameters & \( w_{r}/2\pi \) & \( w_{q}/2\pi \) & \( Q_{i} \) & \( k_{ext} \) & \( \chi \) & \( \alpha \)  & \( \langle T_{1} \rangle \) & \( \langle T_{2R} \rangle \) & \( \langle T_{2E} \rangle \) \\
            \hline
            Qubits & MHz & MHz & \( 10^{4} \) & MHz & KHz  & MHz & \( \mu s \) & \( \mu s \) & \( \mu s \) \\
            \hline
            \( Q_{1} \)  & 9997.4 & 4888.2 & 11.7 & 2.6 & -200.0 & -196.6 & 126 ± 18 & 107 ± 12 & 124 ± 23 \\
            \( Q_{2} \)  & 9386.0 & 4795.6 & 11.8 & 1.3 & -225.0 & -197.2 & 89 ± 13  & 56 ± 15  & 86 ± 12 \\
            \( Q_{3} \)  & 9299.2 & 4807.5 & 6.5  & 1.8 & -200.0 & -196.2 & 61 ± 6   & 44 ± 5   & 102 ± 18 \\
            \( Q_{4} \)  & 8649.5 & 4809.8 & 5.3  & 2.8 & -200.0 & -198.6 & 54 ± 9   & 39 ± 14  & 97 ± 20 \\
            \( Q_{5} \)  & 8755.6 & 4855.3 & 3.1  & 0.8 & -225.0 & -196.4 & 68 ± 10  & 38 ± 4   & 45 ± 10 \\
            \( Q_{6} \)  & 9220.6 & 4824.8 & 6.4  & 0.5 & -225.0 & -194.0 & 63 ± 7   & 49 ± 4   & 68 ± 10 \\
            \( Q_{7} \)  & 9474.2 & 4928.5 & 10.8 & 1.7 & -175.0 & -195.6 & 77 ± 12  & 45 ± 12  & 82 ± 15 \\
            \( Q_{8} \)  & 9908.6 & 4829.5 & 5.8  & 2.0 & -175.0 & -197.2 & 63 ± 8   & 32 ± 7   & 71 ± 8 \\
            \( Q_{9} \)  & 9802.3 & 4963.4 & 7.8  & 4.4 & -250.0 & -195.0 & 24 ± 7   & 24 ± 5   & 33 ± 9 \\
            \( Q_{10} \) & 9535.4 & 4817.2 & 7.3  & 3.2 & -200.0 & -196.9 & 51 ± 10  & 35 ± 9   & 63 ± 17 \\
            \( Q_{11} \) & 9112.8 & 4835.4 & 16.5 & 1.4 & -175.0 & -196.1 & 74 ± 7   & 49 ± 3   & 78 ± 13 \\
            \( Q_{12} \) & 8851.8 & 4777.3 & 11.4 & 0.9 & -250.0 & -196.4 & 92 ± 24  & 57 ± 11  & 76 ± 15 \\
            \( Q_{13} \) & 8943.2 & 4884.0 & 16.7 & 1.6 & -175.0 & -195.3 & 102 ± 13 & 63 ± 7   & 107 ± 22 \\
            \( Q_{14} \) & 9025.3 & 4855.2 & 9.3  & 1.4 & -175.0 & -197.0 & 56 ± 12  & 56 ± 10  & 60 ± 15 \\
            \( Q_{15} \) & 9645.5 & 5040.2 & 7.3  & 1.7 & -225.0 & -196.1 & 55 ± 7   & 58 ± 9   & 65 ± 9 \\
            \( Q_{16} \) & 9728.7 & 4792.8 & 24.6 & 2.7 & -175.0 & -197.5 & 60 ± 6   & 46 ± 4   & 65 ± 9 \\
            \hline
            Statistics  \\
            \hline
            \( \text{Max} \) & 9997.4 & 5040.2 & 24.6 & 4.4 & -175.0 & -194.0 & 126 & 107 & 124 \\
            \( \text{Min} \) & 8649.5 & 4777.3 & 3.1  & 0.5 & -250.0 & -198.6 & 41 & 32 & 45 \\
            \( \mu \) (Mean) & 9333.5 & 4856.5 & 10.1 & 1.9 & -203.1 & -196.4 & 71 & 51 & 78 \\
            \( \sigma \) (Std. Dev)   & 407.12 & - & 5.3 & 0.96  & 26.3 & 1.1 & 21 & 17 & 20 \\
            \( \sigma/\sqrt{N} \) (\( N=16 \)) & 101.8 & - & 1.3 & 0.2 & 6.6 & 0.3 & 5 & 4 & 5 \\
            \hline
        \end{tabular}
    }
    \label{basic_device_parameters}
\end{table}

\subsection*{Single-Qubit Gate Calibration}
\begin{figure}[H]
  \centering
   \includegraphics[width=1.0\textwidth]{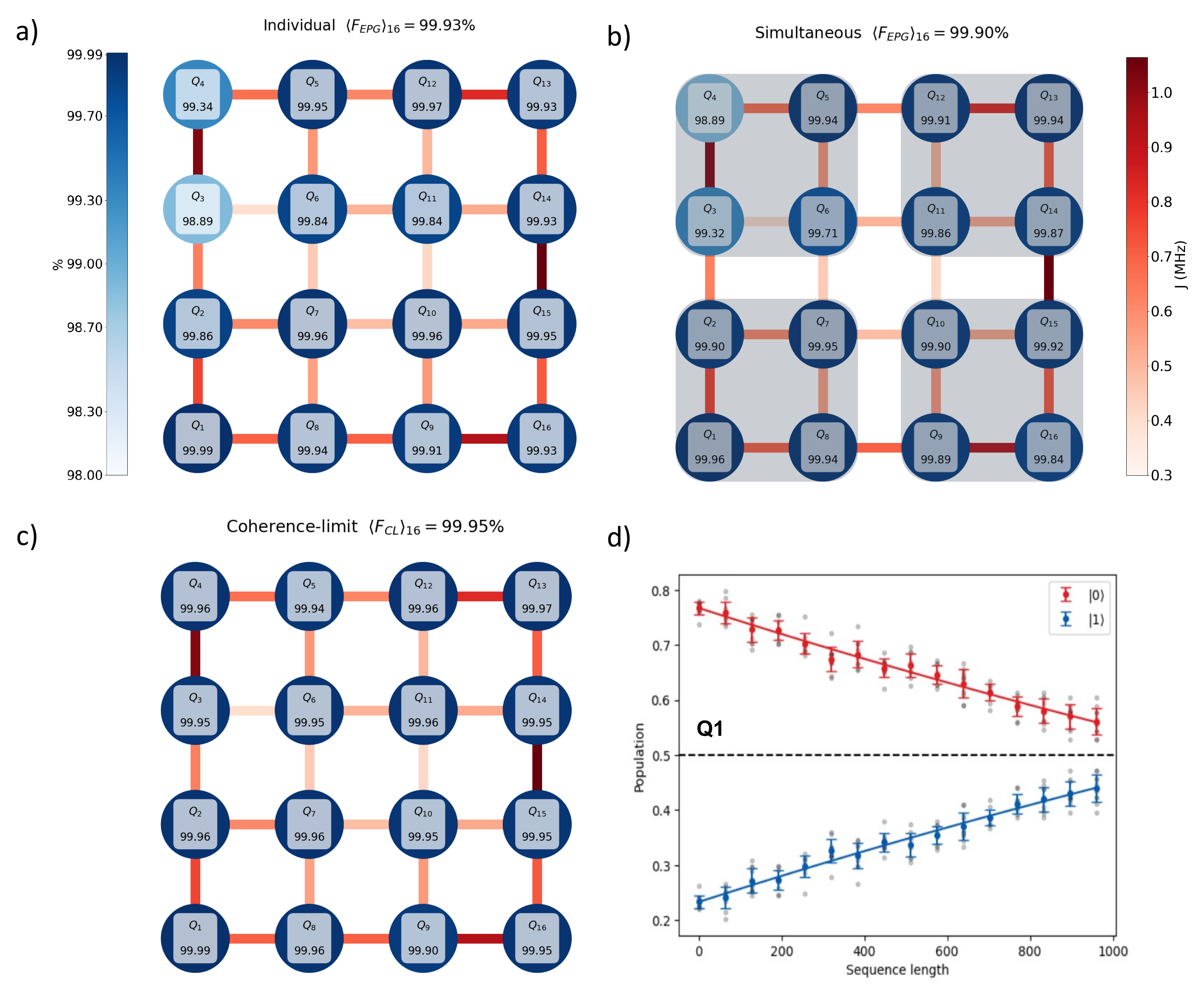}
  \caption{ \textbf{Single-qubit gate fidelities}. (a) Individual and (b) simultaneous error-per physical gate fidelities $F_{EPG}$ with median values across the device compared to coherence-limit physical gate fidelities $F_{CL}$ in (c). Measurements of each gate fidelity is done by randomized benchmarking on $XY$ Clifford decomposition for both individual and simultaneous measurements across the lattice with example of RB measured trace on $Q_{1}$ shown in (d). The coupling across the lattice is shown with the colored bar representing the strength between each qubit.} 
  \label{F5_F_2}
\end{figure}

Randomized benchmarking (RB) ~\cite{chow2009randomized, gambetta2012characterization} experiments were conducted using an \( XY \)-Clifford decomposition for both individual and simultaneous four-qubits gate fidelities. Detailed single-qubit gate fidelities are given in Table. \ref{individual_RB} and Table. \ref{simultaneous_RB}, and shown in Fig.\ref{F5_F_2} with an example of RB measured trace on $Q_{1}$ shown in in Fig.\ref{F5_F_2}(d).

\begin{table}[!ht]
    \centering
    \caption{Individual qubits RB results}
    \resizebox{1.0\textwidth}{!}{ 
        \begin{tabular}{|c|c|c|c|c|c|c|c|c|c|}
            \hline
            Parameters & EPC & EPC Error & \( \mathcal{F}_{EPC} \) & EPG & EPG Error & \( \mathcal{F}_{EPG} \) & CLG & CLG Error & \( \mathcal{F}_{CLG} \) \\
            \hline
            Qubits & Error per Clifford Gate & Error & \% & Error per Physical Gate & Error & \% & Coh. Lim. Error per Gate & Error & \% \\
            \hline
            \( Q_{1} \) & 1.212E-04 & 1.167E-04 & 99.988 & 6.640E-05 & 6.393E-05 & 99.993 & 1.111E-04 & 1.730E-05 & 99.989 \\ 
            \( Q_{2} \) & 2.531E-03 & 2.520E-04 & 99.747 & 1.388E-03 & 1.381E-04 & 99.861 & 3.689E-04 & 5.286E-05 & 99.963 \\ 
            \( Q_{3} \) & 2.014E-02 & 1.237E-02 & 97.986 & 1.108E-02 & 6.780E-03 & 98.892 & 5.375E-04 & 6.706E-05 & 99.946 \\ 
            \( Q_{4} \) & 1.205E-02 & 2.777E-03 & 98.795 & 6.622E-03 & 1.522E-03 & 99.338 & 3.915E-04 & 5.258E-05 & 99.961 \\ 
            \( Q_{5} \) & 8.595E-04 & 7.357E-05 & 99.914 & 4.711E-04 & 4.031E-05 & 99.953 & 5.919E-04 & 1.012E-04 & 99.941 \\ 
            \( Q_{6} \) & 2.852E-03 & 2.096E-04 & 99.715 & 1.564E-03 & 1.148E-04 & 99.844 & 4.672E-04 & 4.662E-05 & 99.953 \\ 
            \( Q_{7} \) & 6.773E-04 & 1.818E-04 & 99.932 & 3.712E-04 & 9.961E-05 & 99.963 & 3.739E-04 & 4.903E-05 & 99.963 \\ 
            \( Q_{8} \) & 1.027E-03 & 6.242E-05 & 99.897 & 5.630E-04 & 3.420E-05 & 99.944 & 4.406E-04 & 3.763E-05 & 99.956 \\ 
            \( Q_{9} \) & 1.591E-03 & 8.031E-05 & 99.841 & 8.719E-04 & 4.400E-05 & 99.913 & 1.024E-03 & 2.056E-04 & 99.898 \\ 
            \( Q_{10} \) & 6.621E-04 & 1.169E-04 & 99.934 & 3.629E-04 & 6.406E-05 & 99.964 & 5.138E-04 & 9.399E-05 & 99.949 \\ 
            \( Q_{11} \) & 2.821E-03 & 2.885E-04 & 99.718 & 1.547E-03 & 1.581E-04 & 99.845 & 3.917E-04 & 4.464E-05 & 99.961 \\ 
            \( Q_{12} \) & 6.173E-04 & 9.181E-05 & 99.938 & 3.383E-04 & 5.031E-05 & 99.966 & 3.720E-04 & 5.922E-05 & 99.963 \\ 
            \( Q_{13} \) & 1.299E-03 & 1.224E-04 & 99.870 & 7.122E-04 & 6.706E-05 & 99.929 & 2.850E-04 & 4.043E-05 & 99.971 \\ 
            \( Q_{14} \) & 1.355E-03 & 8.207E-05 & 99.865 & 7.426E-04 & 4.497E-05 & 99.926 & 5.122E-04 & 9.179E-05 & 99.949 \\ 
            \( Q_{15} \) & 9.704E-04 & 7.331E-05 & 99.903 & 5.318E-04 & 4.017E-05 & 99.947 & 4.898E-04 & 4.853E-05 & 99.951 \\ 
            \( Q_{16} \) & 1.353E-03 & 1.367E-04 & 99.865 & 7.419E-04 & 7.490E-05 & 99.926 & 4.746E-04 & 4.579E-05 & 99.953 \\ 
            \hline
        \end{tabular}
    }
    \label{individual_RB}
\end{table}

\begin{table}[!ht]
    \centering
    \caption{Simultaneous four-qubit sets RB results}
    \resizebox{1.0\textwidth}{!}{ 
        \begin{tabular}{|c|c|c|c|c|c|c|c|c|c|}
            \hline
            Parameters & EPC & EPC Error & \( \mathcal{F} \) & EPG  & EPG Error & \( \mathcal{F} \) & CLG & CLG Error & \( \mathcal{F} \) \\
            \hline
            Qubits & Error per Clifford Gate & Error & \% & Error per Physical Gate & Error & \% & Coh. Lim. Error per Gate & Error & \% \\
            \hline
            \( Q_{1} \) & 7.152E-04 & 1.149E-04 & 99.928 & 3.919E-04 & 6.299E-05 & 99.961 & 1.111E-04 & 1.730E-05 & 99.989 \\ 
            \( Q_{2} \) & 1.782E-03 & 3.456E-04 & 99.822 & 9.768E-04 & 1.894E-04 & 99.902 & 3.689E-04 & 5.286E-05 & 99.963 \\ 
            \( Q_{3} \) & 1.233E-02 & 2.312E-03 & 98.767 & 6.775E-03 & 1.267E-03 & 99.323 & 5.375E-04 & 6.706E-05 & 99.946 \\ 
            \( Q_{4} \) & 2.024E-02 & 7.128E-03 & 97.976 & 1.114E-02 & 3.906E-03 & 98.886 & 3.915E-04 & 5.258E-05 & 99.961 \\ 
            \( Q_{5} \) & 1.137E-03 & 7.717E-05 & 99.886 & 6.232E-04 & 4.228E-05 & 99.938 & 5.919E-04 & 1.012E-04 & 99.941 \\ 
            \( Q_{6} \) & 5.231E-03 & 6.081E-04 & 99.477 & 2.870E-03 & 3.332E-04 & 99.713 & 4.672E-04 & 4.662E-05 & 99.953 \\ 
            \( Q_{7} \) & 9.912E-04 & 1.300E-04 & 99.901 & 5.432E-04 & 7.122E-05 & 99.946 & 3.739E-04 & 4.903E-05 & 99.963 \\ 
            \( Q_{8} \) & 1.170E-03 & 1.065E-04 & 99.883 & 6.412E-04 & 5.836E-05 & 99.936 & 4.406E-04 & 3.763E-05 & 99.956 \\ 
            \( Q_{9} \) & 1.909E-03 & 1.267E-04 & 99.809 & 1.047E-03 & 6.943E-05 & 99.895 & 1.024E-03 & 2.056E-04 & 99.898 \\ 
            \( Q_{10} \) & 1.802E-03 & 2.894E-04 & 99.820 & 9.879E-04 & 1.586E-04 & 99.901 & 5.138E-04 & 9.399E-05 & 99.949 \\ 
            \( Q_{11} \) & 2.529E-03 & 1.516E-04 & 99.747 & 1.387E-03 & 8.307E-05 & 99.861 & 3.917E-04 & 4.464E-05 & 99.961 \\ 
            \( Q_{12} \) & 1.565E-03 & 1.710E-04 & 99.844 & 8.578E-04 & 9.369E-05 & 99.914 & 3.720E-04 & 5.922E-05 & 99.963 \\ 
            \( Q_{13} \) & 1.003E-03 & 9.206E-05 & 99.900 & 5.499E-04 & 5.044E-05 & 99.945 & 2.850E-04 & 4.043E-05 & 99.971 \\ 
            \( Q_{14} \) & 2.383E-03 & 1.273E-04 & 99.762 & 1.307E-03 & 6.974E-05 & 99.869 & 5.122E-04 & 9.179E-05 & 99.949 \\ 
            \( Q_{15} \) & 1.451E-03 & 9.408E-05 & 99.855 & 7.955E-04 & 5.155E-05 & 99.920 & 4.898E-04 & 4.853E-05 & 99.951 \\ 
            \( Q_{16} \) & 2.848E-03 & 3.135E-04 & 99.715 & 1.561E-03 & 1.718E-04 & 99.844 & 4.746E-04 & 4.579E-05 & 99.953 \\ 
            \hline
        \end{tabular}
    }
    \label{simultaneous_RB}
\end{table}

\newpage
\subsection*{Two-Qubit Interaction and Gate Calibration}
The calibration of a CZ gate based on the siZZle interaction, it requires optimizing the drive parameters such that the total ZZ-induced phase accumulation during the gate operation equals $\pi/4$. This involves first tuning up the frequencies of the off-resonant drives to achieve optimal detunings $\Delta_{0,d}$ and $\Delta_{1,d}$ from the qubits' transitions, selecting both optimal amplitudes ($\Omega_0$ and $\Omega_1$) and phase difference ($\Delta = \phi_0 - \phi_1)$ between the off-resonant drives to maximize the $\tilde{\nu}_{ZZ}$ rate, and finally working the gate duration out of the optimal $\tilde{\nu}_{ZZ}$ rate. 

\begin{figure}[H]
  \centering
   \includegraphics[width=1\textwidth]{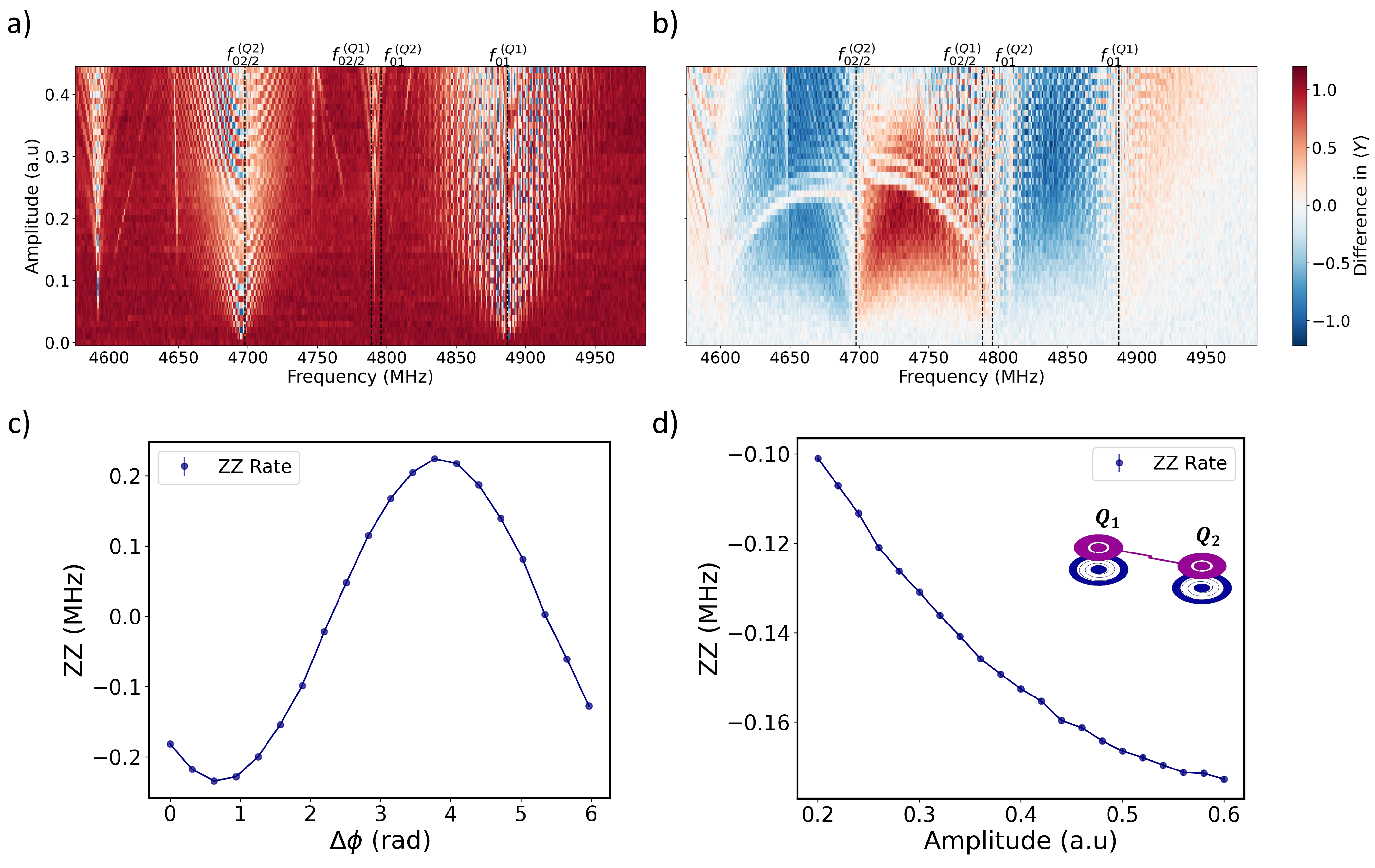}
  \caption{\textbf{Two-qubit interactions and CZ gate Calibration}. 
       (a) Two-dimensional parameter sweep of Stark drive frequency and amplitude, showing the response of the control qubit used to infer interaction stability. (b) Corresponding measurement on the target qubit, capturing the differential phase accumulation due to ZZ interaction. These scans are used to map out the interaction landscape and identify regions of coherent dynamics. (c) Measured modulation of the ZZ interaction rate as a function of the relative phase difference $\Delta = \phi_0 - \phi_1$ between the two off-resonant drives. (d) ZZ interaction strength as a function of equal-amplitude off-resonant drives applied to both qubits, used to extract the optimal amplitude.}
  \label{F8}
\end{figure}

The calibration procedure begins by selecting an initial set of Stark drive parameters. Specifically, a drive frequency and amplitude are first selected to observe an initail ZZ interctaion. This assessment is carried out using Hamiltonian tomography followed by repeated gate tomography, as illustrated in Fig. ~\ref{F9}(b) and (c), respectively. Non-optimal parameters typically lead to unstable or noisy oscillations in the expectation values. In such cases, the parameters are iteratively adjusted until clean, stable oscillations are observed. Once stability is achieved, an automated fitting routine is used to extract the optimal gate duration to proceed to the calibration of a CZ entangling gate.

To tune the ZZ interaction more precisely, we perform a two-dimensional parameter sweep over the Stark drive frequency and amplitude shown in Fig. ~\ref{F8}, building a coarse map of the ZZ interaction rates. Although directly measuring the ZZ rate at every point would be ideal, it is experimentally very expensive. Instead, we fix the Stark pulse duration at $1~\mu$s and use the Hamiltonian tomography sequence in Fig. ~\ref{F9}(b), recording the differential phase accumulation on the target qubit on $\langle Y \rangle$ basis when the control qubit is initialized in either the ground or excited state. This entire mapping procedure takes roughly 12 hours and yields the background for the interaction landscape over a wide range of parameters. Following this, we analyze both the control qubit (in Fig. ~\ref{F8}(a)) and target qubit (in Fig. ~\ref{F8}(b)) dynamics to identify the optimal parameters for high-contrast, coherent interactions.

\begin{figure}[H]
  \centering
   \includegraphics[width=1\textwidth]{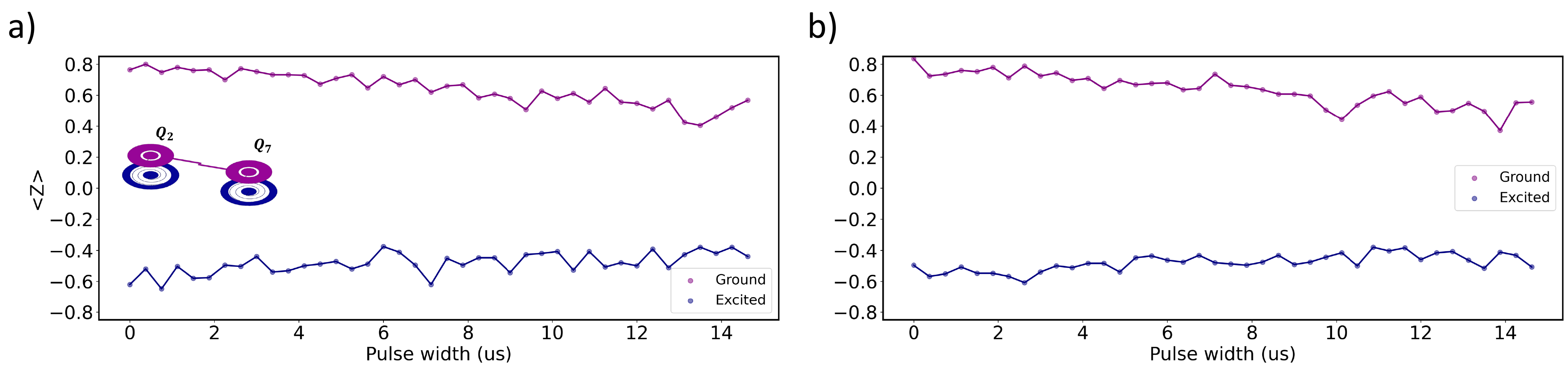}
  \caption{\textbf{Two-qubit interactions and CZ gate Calibration}. 
       States of the control qubit $Q_{2}$ during ZZ-induced phase accumulation on target qubit $Q_{7}$ during  $\langle X \rangle$ measurements in (a) and $\langle Y \rangle$ measurements in (b), for calibrating a CZ gate between $Q_{2}$ and $Q_{7}$. See the corresponding $\langle X \rangle$ and $\langle Y \rangle$ measurements on the target qubit in Fig. ~\ref{F9}(b).}
  \label{F9b}
\end{figure}

\bibliography{Biography}

\end{document}